**Why can't Epidemiology be automated (yet)?**

David Bann[1]*, Ed Lowther[2], Liam Wright[1], Yevgeniya Kovalchuk[2]

[1]Centre for Longitudinal Studies, University College London, London, UK
[2]Centre for Advanced Research Computing, University College London, London, UK
*Corresponding author: david.bann@ucl.ac.uk

Recent advances in artificial intelligence (AI)—particularly generative AI—present new opportunities to accelerate, or even automate, epidemiological research. Unlike disciplines based on physical experimentation, a sizable fraction of Epidemiology relies on secondary data analysis and thus is well-suited for such augmentation. Yet, it remains unclear which specific tasks can benefit from AI interventions or where roadblocks exist. Awareness of current AI capabilities is also mixed. Here, we map the landscape of epidemiological tasks using existing datasets—from literature review to data access, analysis, writing up, and dissemination—and identify where existing AI tools offer efficiency gains. While AI can increase productivity in some areas such as coding and administrative tasks, its utility is constrained by limitations of existing AI models (e.g. hallucinations in literature reviews) and human systems (e.g. barriers to accessing datasets). Through examples of AI-generated epidemiological outputs, including fully AI-generated papers, we demonstrate that recently developed agentic systems can now design and execute epidemiological analysis, albeit to varied quality (see https://github.com/edlowther/automated-epidemiology). Epidemiologists have new opportunities to empirically test and benchmark AI systems; realising the potential of AI will require two-way engagement between epidemiologists and engineers.

**Introduction**

Epidemiology is concerned with understanding the distribution and determinants of health in the population. A sizable fraction of epidemiological research involves secondary data analysis: statistically analysing data collected from cohorts, cross-sectional studies, or other data sources. Such research comprises a series of cognitive tasks currently conducted, or at least overseen, by humans.

Historically, conducting epidemiological research was a slow, manual endeavour: scanning library shelves, reading physical papers, and manually collecting, coding, and analysing data (Figure 1). Technological progress has now led much of this work to be electronic, yet actual scientific progress arguably remains slow despite, for example, the surge in large cohorts and ballooning data volumes—omics, wearables, administrative linkages, etc.—progress in identifying modifiable risk factors for disease has proved elusive.[1][2]

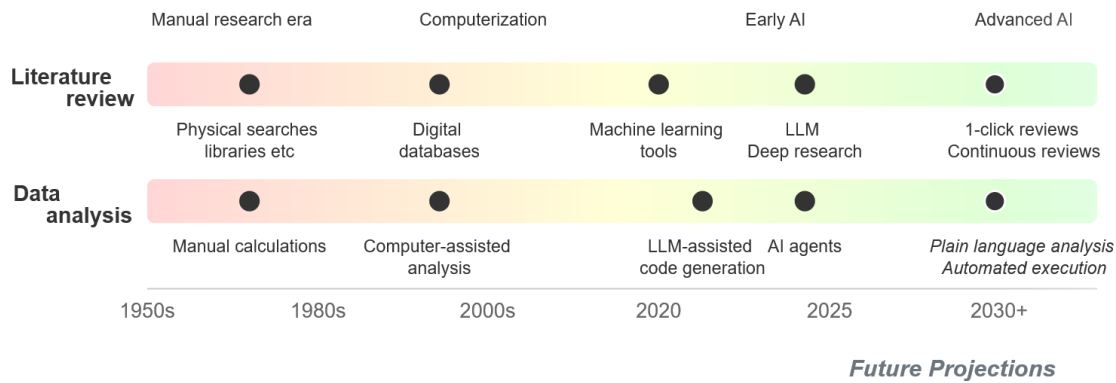

**Figure 1. Technological progress in two key epidemiological tasks (literature reviews and data analysis): from manual work, computerization, to AI augmented research.** Note, in some senses, the final tasks listed (e.g. plain language analysis) are already possible with current AI systems, yet the quality of outputs is of mixed or unknown quality.

AI represents the next step in epidemiology's technological evolution (Figure 1); it can accelerate—or even automate—cognitive tasks, boosting the efficiency of current practice and creating new opportunities for discovery. Epidemiologists often use AI-based tools—sometimes without explicitly knowing it—such as Google Scholar for paper discovery, spell-checkers for writing, and GitHub Copilot for coding.

Despite previous AI 'winters', its current era of development, built around the transformer deep learning architecture[3] that powers modern large language models (LLMs), has generated remarkable progress. LLMs shot into public consciousness in November 2022 with the release of ChatGPT, reportedly the fastest growing consumer product of all time. Many scientists, particularly younger researchers, now use ChatGPT.[4] Other LLMs have since become publicly available and widely used, and billions of dollars invested in their training. LLMs predict the next token (typically, a small piece of text) in a sequence—developed at massive scale, which yields surprisingly useful properties. Productivity increases in tasks relevant to epidemiology have recently been suggested—writing,[5] cognitive tasks,[6] debating/reasoning,[7] and coding.[8]

Here, we map the landscape of epidemiological tasks that rely on existing datasets—from literature review through to idea generation, data access, analysis, write-up and dissemination. We provide a snapshot of available AI tools and a repository containing examples of AI-generated epidemiological output, along with prompt and model details (github.com/edlowther/automated-epidemiology). Finally, we discuss barriers to deeper AI integration and broader implications for the field, including how epidemiologists can contribute to AI development, addressing recent calls.[9] We note that the issues discussed apply to other fields, e.g. social sciences.[10,11]

**Conducting literature reviews**

Systematic reviews generally take one year or more to undertake,[12] with much of this time spent on pain-staking and often painful screening—manually removing the vast majority of irrelevant articles from the selected pool by comparing them against the same inclusion/exclusion criteria. On face value, this breaks the software engineering principle to "automate repetitive tasks", but an additional motivation for automation is to reduce errors: humans do not screen without error.[13]

In recent years, machine learning tools have become available to speed up screening: authors manually screen a smaller subset of abstracts to train models, which then automatically screen the remainder.[14] Such tools appear to increase efficiency,[15,16] widening the scope to produce or update reviews more rapidly (possibly continuously), and undertake more ambitious reviews. LLMs can help in other review-related tasks such as creating synonym/search terms lists and extracting data. Could the entire process of reviewing be automated? Using an agentic AI system (Otto-SR), a 2025 study claimed to have reproduced and updated a Cochrane issue in two days—the equivalent of 12 work-years of traditional systematic review work (assuming one year per review).[17]

For more ad-hoc literature searches, systematic reviews are typically prohibitively costly, e.g. when informing introductions or discussion sections in original research articles. Researchers are increasingly using AI-augmented search tools such as Google Scholar to undertake literature searchers—unlike PubMed, it indexes non-health publications (e.g. economics articles), as well as grey literature. Nevertheless, either tool requires conversion of the search question (e.g. "what effect does X have on Y?") into terms more likely effective for such databases (e.g. "associations between X and Y", an "RCT of X and Y", etc.), in addition to a continued manual search of 'cited by' articles.

LLMs can answer such single questions directly, and recent reasoning LLMs enable AI 'sense checks' before a response is produced. Hallucinations—which raised considerable concern in early models—appear to have reduced.[18] 'DeepResearch' capabilities, made available in several leading LLM tools in recent months, enable more extended searches of scholarly literature; users can check the sources provided via links to the full text.

The generality of LLMs means they can usefully sift, connect, and summarise evidence from far-flung disciplines—a task that otherwise has become progressively harder as scientific output has surged.[19] For epidemiologists, such sources range from mechanistic studies in cells, animal models, human autopsy studies, and the social sciences (e.g. psychology, economics, sociology). AI tools could thus make cross-disciplinary triangulation more feasible.

Hallucinations remain a barrier to the trustworthiness of LLMs, but a human barrier is in accessing research articles. Since the 1970s, the five largest for-profit publishers have steadily increased their market share, accounting for more than half of all papers by 2013.[20] Papers—and even their abstracts—are copyrighted. This creates a particular barrier for open-source AI systems.[21]

Partial access to research articles and the capacity of LLMs to provide highly compelling narratives mean LLMs may mislead.[22] Ongoing evaluation of such systems is required: empirical study of their sensitivity and specificity in search, for example. This is a challenge given their rapid development—closed-source frontier LLMs can be rapidly updated or decommissioned.

**Creativity and generating hypotheses**

It is often assumed that AI systems (particularly LLMs) simply interpolate between data points contained within their training set and are thus not capable of being creative or generating novel ideas—i.e. they are 'stochastic parrots'.[23] Setting aside the 'incremental' nature of modern science, such claims are at least partly empirically testable: an emerging literature suggests that the creative capability of frontier models may match that of humans in discrete small-scale creative tasks.[24] Their abilities in real-life scientific creativity remains uncertain, as does the comparisons of humans alone versus human–AI collaborations in 1) forming hypotheses which advance epidemiology or 2) selecting hypotheses which are tractable and falsifiable[25] given existing data. In other disciplines, such as drug discovery, new scientific findings are seemingly being discovered via AI systems.[26]

We prompted a recently developed AI tool (the AI Scientist[27]) to suggest novel hypotheses across two topics: 1) the links between birth weight and subsequent BMI, and 2) social inequalities in mental health; see github.com/edlowther/automated-epidemiology. Many hypotheses appear to have face validity, e.g. suggesting generally underutilised approaches to causal inference (sibling comparison studies and natural experiments). We note that such suggestions were created in 'one-shot' and are thus the equivalent of a human's first draft. Even if only a fraction of AI-suggested hypotheses are promising, the number that can be created quickly is large and may be especially valuable with humans 'in the loop' to select them: an AI-augmented process akin to human brainstorming.

**Identifying and accessing data**

A common approach in epidemiological research is that groups running specific epidemiological studies (e.g. cohorts or health surveys) publish research focused on using that specific dataset. In this scenario, multiple publications in the literature from different research groups address the same question; yet, subsequently synthesising such evidence (e.g. via meta-analysis) is not always possible due to methodological differences. Consortia integrating multiple studies are one manual approach to circumvent this, but they are typically setup for specific research questions and are hard to maintain in the long run; when their funding ends, they may cease to operate.

AI may enable a bolder default for epidemiological research enabling us to answer, for each research question, what are the possible available datasets that could contribute evidence? Of these, which have harmonizable data? And what does that evidence collectively show?

Current barriers to this include high fixed costs of becoming familiar with datasets, and the fragmented approaches to data discovery and access. Platforms to aid cohort discovery, e.g. the recent Atlas of Longitudinal Data (https://atlaslongitudinaldatasets.ac.uk), are a step forwards in helping identify datasets; yet, using them highlights our barriers: 10 different cohorts may involve 10 separate access systems, with considerable overlap in information requested.

The challenge for data providers is whether a single point of entry can be provided—a cohesive streamlined data access system with interoperable data and necessary safeguards. ORCID provides a centralised and broadly accepted system for verifying researcher identity—could existing centralised systems for data documentation and access be expanded (e.g. UK Data Service for UK cohorts; or the Gateway to Global Aging Data, for older adults) or newly created? Within such systems, AI tools can also facilitate the historically slow and manual process of harmonising data across different datasets[28] (e.g. the Harmony tool[29]).

Finally, AI tools can aid in the creation of new epidemiological data. In existing cohorts for example, data held in historic non-electronic form (e.g. paper questionnaires or microfiche) can be digitised using automated optical character recognition (OCR) tools. Such tools can also be used to create new retrospective cohorts: many hundreds of papers have now cited the cohort profiles that arose from the discovery and digitisation of records, which formed the basis for the Hertfordshire[30] and Lothian cohort studies.[31] AI tools could also improve existing metadata (e.g. annotating questionnaires with associated variable names).

**Analysing data**

Much like in literature reviews, epidemiologists are increasingly augmented by AI when analysing data. Rather than manually typing out each letter when coding, AI autocomplete such as GitHub Copilot can speed up code writing. Frontier LLMs are now able to create a complete draft of code in response to a plain-language prompt and then execute this code. The promise is that rapid, autonomous generation of research code will enable human researchers to spend more time at higher levels of abstraction, e.g. thinking carefully about designing research strategies.

We prompted an agentic AI framework ([Data Analysis Crow](#)) to address two research questions and provided simulated data. The responses yielded an analytical plan, analytical code, execution of this code, and visualisations—see [github.com/edlowther/automated-epidemiology](#) for full workbooks and Table 1 for a summary. While the outputs were in our view impressive, they did contain errors, and in some cases failed entirely depending on the underlying LLM used. This suggests 1) the choice of LLM seems important and 2) code review remains essential.

Often epidemiologists specialise in one piece of software or programming language (e.g. SAS, SPSS, Stata, R). In one sense, specialisation is increasingly not needed as the barrier to entry lowers to code in multiple languages. What will remain important is the clear articulation of the goals in plain language and code review. The fact that AI systems provide analytical syntax also aids reproducibility: something that fewer than 2% of health researchers currently do.[32]

The more mundane aspects of data analysis could also be accelerated by AI. Data cleaning for example, is often a highly manual and time-intensive task required even for well-used datasets, leading to considerable duplication of work. Assuming data cleaning involves 1 month of unnecessary work (cleaning data, which should have otherwise been centrally cleaned)—a task ordinarily repeated across 1,000 papers—1,000 months (83 years) of scientists' time could be saved in future. A cursory look at the literature suggests at least six distinct AI data cleaning tools from 2024 onwards that claim varying levels of accuracy in data cleaning.[33-38] Whether such tools are useful in epidemiological applications remains to be seen. A challenge for epidemiologists will be to make sense of the bewildering numbers of tools released in AI-related fields: the curation of epidemiological benchmarks could provide objective criteria by which they can be continually evaluated.

Barriers to the use of AI in data analysis include the current frequent need for uploading data to cloud providers: this is not possible for many health-related datasets held in sandboxed secure computing environments. Researchers could instead use local open-source models—such models have historically been weaker than the closed-source models, yet in recent months, the gap has considerably narrowed.[39] Alternatively, the AI tools can be restricted to accessing metadata (e.g. variable names, labels, and result output) rather than the raw data, or data owners could release synthetic versions of their data.

Epidemiologists will, as ever, need to balance two sets of competing risks. The first is a risk of high-profile data leaks if AI tools are used irresponsibly; the second is a risk that scientific discovery is restricted if such tools are not used. The later risk is often overlooked in our view, despite threats to the continued existence of epidemiological studies (e.g. funding uncertainty and declining response rates).

**Writing up**

Remarkably, given a simple plain-language prompt, frontier LLMs can produce entire epidemiological research papers; [github.com/edlowther/automated-epidemiology](#) shows examples of this using multiple LLMs, each instructed to write a paper on the association between birth weight and adult BMI using a simulated dataset we provided.

Where do such papers sit in the current distribution of human-created epidemiological research papers? Despite being given very little contextual background, the highest quality amongst our AI-created outputs (produced by ChatGPT's o3 model) on face value appeared to satisfy many commonly used consensus criteria for reporting of epidemiology studies (STROBE guidelines[40]). It also showed signs of reasoning: it identified a sex interaction (associations differed in direction by sex) that we had introduced into the simulated data, despite the prompt instructing the LLM to analyse "adjusted by sex."

In other respects, the AI-produced papers are of low quality, e.g. incorrect referencing and the omission of result items. Yet, as demonstrated in the "Analysing data" section, LLMs can (with tool calling) produce compelling figures and tables. Thus, current barriers to producing high-quality, AI-generated manuscripts may partly reflect limitations in how effectively the model is prompted or integrated with tools.

**Full (end-to-end) automation**

Full automation of epidemiological research papers—from generating the idea all the way through to write-up—is a logical consequence of the capability of AI in each component chained together. Such tools as AI Scientist[27] and data-to-paper[41] are recent open-source examples of this. A fruitful avenue of future research is the evaluation of such systems tailored for epidemiology, e.g. relative to human-generated and AI-with-human-in-the-loop-generated papers.

**Dissemination**

Researchers are increasingly encouraged to share their research with non-specialist audiences such as the public and government policymakers, and more generally engage in continued public discourse. From the perspective of scientists already struggling with a 'mountain of small things',[42] such tasks may be unwelcome—yet, if public discourse is dominated by a small, vocal, and unrepresentative minority of scientists, evidence-based policy may suffer.[43] LLMs can speed-up the creation of blogs, lay summaries, and social media content if provided prompts and context (e.g. the research paper). Entire podcasts can now be entirely automated; an AI-generated podcast based on this article can be found in github.com/edlowther/automated-epidemiology. If AI increases efficiency, researchers may be able to move towards a deeper engagement with evidence-based policy—rather than simply advocating that their own work should change policy, creating unbiased evidence across the entire evidence landscape for example, and carefully considering policy trade-offs.[44]

**Overall utility**

In our judgement, AI's current capability suggests a promising future to accelerate epidemiology. This is the case whether AI is used for narrow tasks under close human supervision; as a research assistant or collaborator[45][46] with human oversight; as an expert; or—more controversially—as a semi- or fully-autonomous research agent.[47] Each may bring benefits to epidemiology, with further integration an evolving combination of both human-system and AI-capability barriers.

A promise of AI in the short to medium term is that it could enable more time to be spent on high-level tasks (e.g. designing new research questions or data collections) rather than low-level tasks that are often undesired (e.g. repetitive admin tasks) or uninspiring (e.g. writing code to recode variables); Figure 2. The blend is at our discretion: some investment in low-level tasks is likely helpful or even necessary to learn (e.g. to deeply understand data, to comprehend statistical methods). If the quality rather than quantity of our outputs is incentivised, the net result could be higher quality science and a bolstering of our discipline. We live in an era that incentivises scientists to produce masses of papers of questionable quality, including their direct purchase online (https://www.youtube.com/watch?v=3Yb60j6haZ8&t=2689); this is a human, not an AI problem.

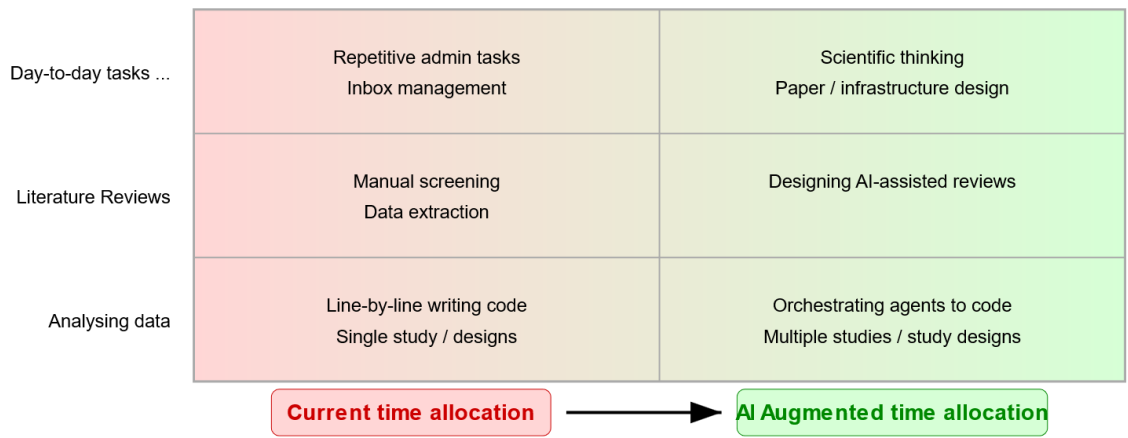

**Figure 2. Could AI help to liberate epidemiologists to focus on higher-level tasks? Simplified illustration of our suboptimal current time allocation (left) versus idealized AI-augmented allocation (right).**

*Existential risks?*

AI developments are rapidly lowering the costs of cognitive tasks and may ultimately lower demand for human epidemiologists; particularly junior epidemiologists, who traditionally lead on writing and analysis tasks, overseen by a senior colleague. If unchecked, this trend could damage the training pipeline, leading to fewer epidemiologists across all levels and thus a collapse in the discipline.

Scientific careers are already uncertain, with rates of pay for epidemiologists generally lower than other technical sectors (e.g. tech/pharma/finance). Will our brightest minds wish to become epidemiologists in future? Addressing structural problems (pay, security) is one route to attract talent. Another is the appeal of working on interesting and important problems—the integration of AI with epidemiology is one. For example, can AI accelerate or automate epidemiology? Can AI benchmarks be tailored/newly created for epidemiology? What biases and risks can AI systems introduce? Can the methods proposed in the AI literature be used to improve prediction[48] and inference[49] in epidemiology? How will AI influence population health?

AI could assist, augment, automate aspects of epidemiology in the future. If AI in its current iteration was to take over human intelligence entirely, our existential role could be temporary: to produce new 'tokens' (data, papers), which vast multi-billion-dollar companies use to train AI systems without our consent. Whether such scenarios are good or bad for scientific discovery or humanity at large remains an open question, which epidemiologists can and should contribute to. Realising the potential of AI will require two-way engagement between epidemiologists and engineers.

**Funding**

DB and LW are funded by the Economic and Social Research Council (ES/W013142/1).

**Contributions**

Wrote the first draft: David Bann. Conducted analysis: Ed Lowther, David Bann. All authors contributed to generating ideas, compiling material, reviewing and revising the text, as well as the accompanying repository.

**Table 1: Evaluating AI-generated analysis: results from the Data Analysis Crow**

| | | Large language model evaluated | |
|---|---|---|---|
| Association | Challenge | GPT-4.1 | Claude Sonnet 4 |
| **Birth weight → BMI** | Data cleaning | ✅ Derived BMI correctly<br>✅ Removed implausible values | ✅ Derived BMI correctly<br>⚠️ Removed some but not all implausible values |
| | Designing and executing analytical plan | ✅ Created analytical plan<br>⚠️ Errors (e.g. <1.2m cases excluded not <1m)<br>✅ Executed results: Tables/Figures. | ✅ Created analytical plan<br>⚠️ Partial results (API crash) |
| | Analysis outcome | ✅ Correct interpretation | ⚠️ Analysis incomplete |
| **Income → mental health** | Data cleaning | ⚠️ Identified sex, assumed value-labels<br>✅ Log-transformed income | ⚠️ Identified sex, assumed value-labels<br>✅ Rescaled income |
| | Designing and executing analytical plan | ✅ Created analytical plan<br>✅ Executed results: Tables/Figures. | ✅ Created analytical plan<br>⚠️ Partial results (API crash) |
| | Analysis outcome | ✅ Correct interpretation | ⚠️ Analysis incomplete |

Notes: Each item was evaluated as follows: ✅: correct or plausible result; ⚠️: error or concern identified. BMI (body mass index), API (Application Programming Interface). A simulated dataset was provided, available on the accompanying repository: https://github.com/edlowther/automated-epidemiology. The Data Analysis Crow is available at https://github.com/Future-House/data-analysis-crow.